\documentstyle[prl,aps,graphicx]{revtex}
\newcommand{\be}{\begin{equation}}
\newcommand{\ee}{\end{equation}}
\newcommand{\ba}{\begin{array}{l}}
\newcommand{\ea}{\end{array}}

\begin{document} \draft %\date{\today} %\large 
\title{Gravitational energy of simple bodies:\\
the method of negative density} \author{Zakir F. Seidov}
\address{\it Department of Physics, Ben-Gurion University,
Beer-Sheva, 84105, Israel\\email:seidov@bgumail.bgu.ac.il}
\maketitle \section*{} \subsection*{Introduction}
In Newtonian theory of gravitation, there are suprisingly  few exact
solutions for the proper  energy of the homogeneous self-gravitating
bodies: a) ellipsoid  (including sphere as a particular case) \cite{Dub},
b) concave bispherical lens (including spherical segment as a particular
case) \cite{KA} (KA), and c) rectangular parallelepiped (including  cube
as a particular case) \cite{ABC}. We present here a method to  obtain the
formulas for potential energy of  homogeneous bodies. The method includes
a notion of {\it negative density}.  Here we use this method, first, to
check the formulas of KA and, second, to get new formula for the
homogeneous concavo-convex lens. To this end, we widely  used Mathematica
\cite{Wolf}. \subsection*{Homogeneous sphere and segment}
Here we use  the method of negative density (MND) to check the  formula by
KA for potential energy of homogeneous spherical segment.
 \begin{figure}  \label{SEGfig} \includegraphics[scale=.4]{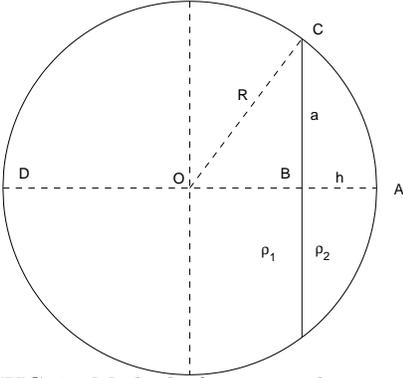}
 \caption{Method of negative density. We put into homogeneous  sphere, of
radius $R=OA=OC=OD$ and density $\rho_1$, a spherical segment
of the same radius $R$, of height $h=AB$ and radius of base $a=BC$.
As a result, we have inhomogeneous spherical body comprised of two 
homogeneous segments with the same radius and base, $R$ and $a$,
 respectively. One segment has density  $\rho_1+\rho_2$ and height $h$,
the other segment has  density $\rho_1$  and height  $2\,R-h=BD$.
}\end{figure} In Fig. 1 we show the setting of the problem. First, we have
the homogeneous sphere with radius $R=OA=OC=OD$ and density $\rho_1$.
Gravitational potential, $\varphi$, inside the homogeneous sphere, as 
function of radius $r$ is:\be\varphi=-2\,\pi\,G\,\rho_1\,(R^2-r^2/3);\ee
and the gravitational energy  of the sphere is:
\be W_{sphere}={1\over 2}\int\limits_M \varphi\,dm\,=-{16\over
15}\pi^2G\,R^5\rho_1^2.\ee Now we put into the sphere  the additional
spherical segment  with density $\rho_2$, the same radius $R$, with height
$h=AB$ and  radius of base $a=BC$. Note that the total matter density in
the region occupied by this segment is $\rho_1+\rho_2$.\\ Potential energy
of interaction of the sphere  and the additional segment is obtained by
integration of $\varphi*\rho_2$  over the volume of the segment:  
\be W_{int}=\int \limits_V \varphi\,\rho_2\,d\,v=
 \int\limits_{R-h}^R\,\varphi\,[ 2\, \pi \,\rho_2\, r\,(r-R +h)]\,dr;\ee
we have: \be W_{int}=-{{G\,{h^2}\,{{\pi }^2}\,(h^3 - 5\,h^2\,R + 20\,R^3 )
\, {\rho_1}\,{\rho_2}}\over {15}}.\ee According to KA, the gravitational
energy of the homogeneous spherical segment with radius $R$, height $h$,
 radius of base $a$ and density $\rho_2$ is ($a=(2 R h -h^2)^{1/2}$):
\be \label{Wseg} W_{segment}(a,h,R,\rho_2)=- {1\over 9}\pi G \rho_2^2
[{3\over 2}\pi\,h^4({h\over 5}-R)+
8\,[3 R^4 a -R^2 a^3-{2\over 5} a^5 -6 R^4 (R-h) \arctan{h\over a}]].\ee
{\it If this formula is correct}, then the gravitational energy of the
homogeneous spherical segment with radius
$R$, height $2\,R-h$, base's radius $a$ and density $\rho_1$ is: 
$\quad\quad W_{ad.seg}=W_{segment}(a,2\,R-h,R,\rho_1).$
We shall show that this is indeed correct.\\
We may look at the body in the Fig. 1 in twofold way:  as \\
a)homogeneous sphere plus homogeneous segment, or as\\ 
b)two homogeneous segments with same 
$a$ and $R$ but with different heights and densities.\\
We can not calculate the total gravitational energy of the body in the
case b) as we do not know an external potential of spherical segment.
 However, we  can calculate the total  gravitational energy of the body in
the case ~a), $W_a$, as a sum of three terms:\\
the proper gravitational energy of homogeneous sphere, Eq. (2)\\
the proper gravitational energy of homogeneous segment, Eq. (5), and \\
the potential energy of {\it interaction} of sphere and segment, Eq. (4).
\be W_a=W_{sphere}+W_{segment}+W_{int}.\ee
Now  the essence of the negative density method: 
{\it we take} $\rho_2=-\rho_1$!\\ Then, turning to the case b), the body
consisted of two segments gives us  one (right) "segment" with {\it zero
density}, and the other (left) homogeneous  segment with "ordinary"
positive density $\rho_1$. In fact, we have only one (left)
segment with parameters $a,\,2\,R-h,\,R,\,\rho_1$. It means that if we
take $\rho_2=-\rho_1$, we  should get , from Eq. (6), the proper potential
energy of "left" segment.\\ As a result, we have the identity for formula
of potential energy of homogeneous spherical segment:
\be \label{ident} W_{segment}(a,2 R - h,R,\rho)
-W_{segment}(a,h,R,\rho)={{G\,{{\pi }^2}\,\left( h - R \right) \,
     {{\left( -{h^2} + 2\,h\,R + 4\,{R^2} \right) }^2}\,
  {{{\rho}}^2}}\over {15}},\ee which is correct if we take into account
the evident relation: 
{$\arctan{h\over a}+\arctan{2\,R-h\over a}={\pi\over 2}$.}
This is a rather rigorous check of the formula (\ref{Wseg}) for potential
energy of spherical segment. We conclude that the method of negative
density is a powerful method of checking the sophisticated formula of
gravitational energy of homogeneous spherical segment.\\
However our MND may be also applied to check the even more complex
formulas by KA for the proper potential energy of symmetric and asymmetric
concave  lenses, and also to obtain the formula for potential energy of
concavo-convex lens.  \subsection*{Homogeneous sphere and asymmetric
convex lens} \begin{figure}  \label{ASLfig}
\includegraphics[scale=.4]{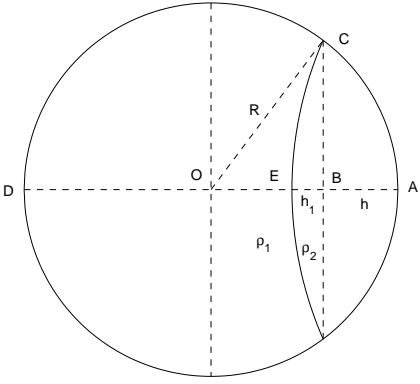}
 \caption{Homogeneous sphere plus asymmetrical lens.}\end{figure}
In Fig. 2 we show the setting of the problem. First, we have the
homogeneous sphere with radius $R=OA=OC=OD$ and density $\rho_1$.
Then, we put into this sphere the  homogeneous, with density $\rho_2$,
asymmetrical bispherical lens, comprised of two homogeneous spherical
segments: one (right) segment  of the same radius $R$, of height $h=AB$
and radius of base $a=BC$; the other (left) segment  of  radius $R_1>R$,
of height $h_1=BE<h=AB$ and the same radius of base $a=BC$.
As a result, we have an inhomogeneous spherical body comprised of two 
homogeneous parts: one asymmetric convex lens (at right) and other 
asymmetric bispherical concavo-convex lens with  surfaces having the
 radii of curvature  $R$ (left surface) and $-R_1$ (right surface).
 The asymmetric convex lens  has density  $\rho_1+\rho_2$ and central
thickness $h+h_1$, the other asymmetric concavo-convex lens has
 density $\rho_1$  and central thickness $h_2=DE=2\,R-h-h_1$.\\
This last figure is of a new kind of homogeneous figures and we are going
to calculate the proper gravitational potential energy of this
 lens using MND {\it provided the  gravitational energy of asymmetric
convex lens is known}. \\ According to KA (see their Eq. (92)) the
gravitational energy of homogeneous asymmetric lens is:
\be \ba d=R_2+R_1-h_2-h_1; a=\sqrt{2\,R_1h_1-h_1^2}=
\sqrt{2\,R_2h_2-h_2^2;}\\ W_{ASL}(\rho_2,a,R_1,R_2)=
-{\pi G \rho_2^2\over 9} \{\pi [a^2 (h_2+h_1)+R_2 h_2^2+
R_1 h_1^2] [R_2^2+R_1^2+R_2 h_2 +R_1 h_1-{h_2^2+h_1^2\over 2}]+\\
 \pi [h_1 (R_1-h_1) (a^2+R_1 h_1)-h _2(R_2-h_2) (a^2+R_2 h_2)] (h_2- 
h_1+R_1-R_2)-\\ \pi [h_2^3 (2 R_2^2-R_2 h_2+{h_2^2\over 5})+ h_1^3 (2
R_1^2-R_1 h_1+{h_1^2\over 5})]-3 \pi (R_2^3 h_2^2 + R_1^3 h_1^2)-\\
 {\pi (R_2^3+R_1^3)\over d}
[h_2^2 (3 R_2 -h_2)+h_1^2 (3 R_1 -h_1)]- {32\over 3} a^5-
8 a^3(R_2^2+R_1^2)-\\ {8 a\over d}[R_2^2 (R_2-h_2) (R_2^2+{2\over 3} a^2)+
R_1^2 (R_1-h_1) (R_1^2+{2\over 3} a^2)]+24 a (R_2^4+R_1^4)+\\
16\, R_2^4 [{R_2^2\over d}-3 (R_2-h_2)] \arctan{h_2\over a}+
16\, R_1^4 [{R_1^2\over d}-3 (R_1-h_1)] \arctan{h_1\over a}\};\ea\ee
here $d$ is distance between two spheres "generating" asymmetric
lens. Potential energy of ASL in the gravitational field of homogeneous
 sphere is (we assume here that  $\rho_2=-\rho_1$): \be \ba
W_{intASL}=\frac{- G\,\pi^2\,\rho_1^2}{90\,d}(R_1 + R_2 - d)^2 \\
\biggl( d^4 + 2\,d^3\,(R_1 + R_2)+ 5\,( R_1 - R_2 )^2\,( 7\,R_1^2 - R_2^2 )-\\
 6\,d^2\,( 2\,R_1^2 -R_1\,R_2 +  2\,R_2^2 )- 2\,d\,( 13\,R_1^3 + 
9\,R_1^2\,R_2 +  9\,R_1\,R_2^2 - 7\,R_2^3)\biggr). \ea \ee
Not repeating the procedure of previous section we present here only the
formula for potential energy of concavo-convex lens in the next compact
form: \be W_{CCL}=W_{sphere}+W_{intASL}-
W_{ASL}(-\rho_1,a,R_1,R_2); \ee see Eqs. (2), (9), (8).
\subsection*{Homogeneous sphere and concavo-convex lens}
\begin{figure}  \label{CCLfig} \includegraphics[scale=.5]{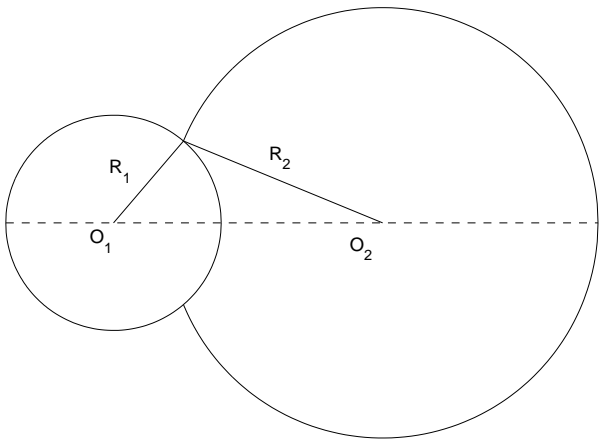}
 \caption{Homogeneous sphere plus concavo-convex lens.}\end{figure}
In this section we calculate the
potential energy of asymmetric convex lens {\it provided the potential 
energy of concavo-convex lens is known}. The setting of the problem is
 shown in Fig. 3. Here we attach to homogeneous sphere (at left) of
radius $R_1$ and density $\rho_1$, the concavo-convex lens with 
radii of surface $R_1$ and $R_2$ and density $\rho_2$. Now we should 
calculate potential energy of CCL in the (external) field of sphere.
We have ($d$ is the distance $O_1O_2$ between centers of spheres, 
{$R_1-R_2\leq\,d\,\leq\,R_1+R_2$}:
\be W_{intCCL}= {{4\,G\,{{\pi }^2}\,{\rho_1}\,{\rho_2}\,
     {{{R_1}}^3}\, \left( {d} - {R_1} - 2\,{R_2} \right) \,
     {{\left( {d} - {R_1} + {R_2} \right) }^2}}\over {9\,{ d}}}.\ee
 Taking $\rho_2=\rho_1$ we get the formula for potential energy of
homogeneous figure in Fig. 3 as a sum of three terms 
(see Eqs. (2), (11), (10)):
\be W_{sphere}+W_{intCCL}+W_{CCL},\ee  which gives the potential energy of
ASL coinciding with $W_{ASL}(\rho_1,a,R_1,R_2)$, see Eq. (8).
\subsection*{Conclusion} In conclusion, we presented here the method of
negative density which is a powerful tool to check the complicated
formulas by KA for potential energy of homogeneous bispheric lenses. The
method allows to obtain new formulas for the potential energy of
homogeneous bodies as well. Though here we used MND only to study
homogeneous bispherical lenses, the MND allows to obtain many new
formulas, e.g. potential energies of spheres (and ellipsoids) and
rectangles with spheres, ellipsoids and rectangles laying within (or
adjacent to) them. The density $\rho_2$ of "additional" bodies may be
different from density $\rho_1$  of the "parent" bodies or, according to
MND, may be equal to $-\rho_1$.  We leave the  study of these so called
"simple" bodies for next communications.   \end{document}